\documentclass[conference]{IEEEtran}
\IEEEoverridecommandlockouts
\usepackage{cite}
\usepackage{amsmath,amssymb,amsfonts}
\usepackage{graphicx}
\usepackage{textcomp}
\usepackage{xcolor}
\usepackage{booktabs}
\usepackage{multirow}
\usepackage{array}
\usepackage{algorithm}
\usepackage{algpseudocode}

\usepackage[utf8]{inputenc}
\usepackage[T1]{fontenc}

\def\BibTeX{{\rm B\kern-.05em{\sc i\kern-.025em b}\kern-.08em
    T\kern-.1667em\lower.7ex\hbox{E}\kern-.125emX}}
\begin{document}

\title{Deep Learning Enhanced Multi-Day Turnover Quantitative Trading Algorithm for Chinese A-Share Market
}

\author{\IEEEauthorblockN{Yimin Du}
\IEEEauthorblockA{
Beijing, China\\
sa613403@mail.ustc.edu}
}

\maketitle

\begin{abstract}
This paper presents a sophisticated multi-day turnover quantitative trading algorithm that integrates advanced deep learning techniques with comprehensive cross-sectional stock prediction for the Chinese A-share market. Our framework combines five interconnected modules: initial stock selection through deep cross-sectional prediction networks, opening signal distribution analysis using mixture models for arbitrage identification, market capitalization and liquidity-based dynamic position sizing, grid-search optimized profit-taking and stop-loss mechanisms, and multi-granularity volatility-based market timing models. The algorithm employs a novel approach to balance capital efficiency with risk management through adaptive holding periods and sophisticated entry/exit timing. Trained on comprehensive A-share data from 2010-2020 and rigorously backtested on 2021-2024 data, our method achieves remarkable performance with 15.2\% annualized returns, maximum drawdown constrained below 5\%, and a Sharpe ratio of 1.87. The strategy demonstrates exceptional scalability by maintaining 50-100 daily positions with a 9-day maximum holding period, incorporating dynamic profit-taking and stop-loss mechanisms that enhance capital turnover efficiency while preserving risk-adjusted returns. Our approach exhibits robust performance across various market regimes while maintaining high capital capacity suitable for institutional deployment.
\end{abstract}


\section{Introduction}

The evolution of quantitative trading in Chinese capital markets has been marked by increasing sophistication in both methodological approaches and technological implementation. The A-share market, characterized by its unique regulatory environment, high retail participation, and distinctive volatility patterns, presents both exceptional opportunities and significant challenges for algorithmic trading strategies. Traditional approaches often fall into two categories: high-frequency strategies with limited capital capacity, and long-term fundamental strategies that may miss intermediate-term market inefficiencies.

This paper addresses these limitations by introducing a comprehensive multi-day turnover quantitative trading algorithm that bridges the gap between short-term tactical allocation and medium-term strategic positioning. Our approach is fundamentally built upon the integration of deep learning techniques with traditional quantitative methods, creating a robust framework that can adapt to the dynamic nature of Chinese equity markets while maintaining consistent risk-adjusted performance.

The core innovation of our methodology lies in its multi-layered approach to market analysis and position management. Rather than relying on a single predictive model or trading signal, we construct a sophisticated ecosystem of interconnected components that work synergistically to identify opportunities, size positions appropriately, and manage risk dynamically. This holistic approach enables our algorithm to capture alpha from multiple sources while maintaining robust risk controls across varying market conditions.

Our framework consists of five integrated modules, each designed to address specific aspects of the trading process. The cross-sectional prediction module employs deep neural networks to identify stocks with superior relative performance potential, moving beyond traditional factor-based approaches to capture complex non-linear relationships in market data. The opening signal distribution analysis component utilizes advanced statistical techniques to identify arbitrage opportunities in opening auction mechanisms, a particularly important feature given the unique characteristics of A-share market microstructure.

The position sizing algorithm represents a significant advancement over conventional approaches by incorporating both market capitalization and liquidity dynamics in real-time, ensuring optimal capital allocation while respecting market impact constraints. Our grid-search optimization framework for profit-taking and stop-loss parameters goes beyond static rule-based approaches, dynamically adapting to market conditions to optimize the trade-off between profit maximization and risk control.

Finally, the multi-granularity market timing component integrates analysis across multiple time horizons and volatility regimes, enabling the algorithm to make nuanced decisions about entry and exit timing that account for broader market dynamics while maintaining focus on individual security selection.

The key contributions of this research include: (1) A novel integration architecture that combines cross-sectional prediction with opening signal distribution analysis for enhanced stock selection and precise entry timing; (2) An adaptive position sizing algorithm that dynamically balances market impact, liquidity constraints, and capital efficiency; (3) A comprehensive grid-search optimization framework that simultaneously optimizes profit-taking, stop-loss, and holding period parameters; (4) A multi-granularity volatility analysis system that incorporates market regime identification for robust timing decisions; (5) Extensive empirical validation demonstrating superior risk-adjusted performance with institutional-scale capital capacity.

\section{Literature Review and Related Work}

The landscape of quantitative trading research has evolved dramatically over the past two decades, with particular acceleration in applications to Asian markets. Traditional factor-based models, pioneered by \cite{fama1993common} and extended by \cite{carhart1997persistence}, established the foundational framework for systematic equity strategies. However, these linear models often fail to capture the complex, non-linear relationships present in modern financial markets, particularly in emerging market contexts such as the Chinese A-share market.

The integration of machine learning techniques into quantitative finance has opened new avenues for alpha generation. Early applications focused on traditional supervised learning approaches, with \cite{khandani2010consumer} demonstrating the effectiveness of support vector machines for stock return prediction. The advent of deep learning has further revolutionized the field, with \cite{fischer2018deep} showing that LSTM networks can effectively capture temporal dependencies in financial time series.

Recent research has increasingly focused on the application of sophisticated neural network architectures to financial prediction tasks. \cite{sezer2020financial} provides a comprehensive survey of deep learning applications in financial forecasting, highlighting the superior performance of neural networks over traditional econometric models in many contexts. \cite{zhou2021informer} introduces the Informer architecture, which addresses the computational challenges of applying transformer models to long sequences commonly encountered in financial data.

Cross-sectional analysis has emerged as a particularly powerful approach for equity selection, as it naturally controls for market-wide factors while focusing on relative performance differences. \cite{harvey2016and} demonstrates that cross-sectional models can significantly improve the information ratio of equity strategies by reducing exposure to systematic risk factors. This approach is particularly relevant in the Chinese market context, where sector rotation and style factors play significant roles in return generation.

The Chinese A-share market presents unique characteristics that distinguish it from developed markets. \cite{carpenter2021anomalies} documents numerous market anomalies specific to Chinese equities, including calendar effects, momentum patterns, and liquidity premiums that differ substantially from those observed in Western markets. \cite{wang2020machine} specifically examines the application of machine learning techniques to Chinese equity selection, finding that ensemble methods combining multiple algorithms tend to outperform single-model approaches.

Volatility modeling has been a central focus of financial econometrics, with \cite{engle2002dynamic} introducing the DCC-GARCH framework for multivariate volatility modeling. More recent work has explored the application of neural networks to volatility prediction, with \cite{ravi2017deep} demonstrating that deep learning models can capture volatility clustering and regime changes more effectively than traditional GARCH models.

Position sizing and portfolio construction represent critical components of any systematic trading strategy. The Kelly criterion, introduced by \cite{kelly1956new} and adapted for financial applications by \cite{thorp2006kelly}, provides a theoretical framework for optimal position sizing under uncertainty. However, practical implementations must account for transaction costs, market impact, and liquidity constraints. \cite{almgren2001optimal} develops a framework for optimal execution that balances market impact with timing risk, while \cite{cartea2015algorithmic} extends this work to high-frequency trading contexts.

Risk management in quantitative strategies has evolved from simple volatility targeting to sophisticated multi-dimensional approaches. \cite{israelsen2005refinement} proposes modifications to the traditional Sharpe ratio that better capture tail risk characteristics. \cite{agarwal2004risks} examines the risk characteristics of systematic trading strategies, emphasizing the importance of drawdown control and regime-aware risk management.

The integration of alternative data sources has become increasingly important in modern quantitative strategies. \cite{kolanovic2017big} provides an overview of big data applications in systematic trading, while \cite{bollen2011twitter} demonstrates the predictive power of sentiment analysis for market movements. However, the regulatory environment in China limits the availability of some alternative data sources, making the optimization of traditional market data particularly important.

Market timing, long considered a challenging aspect of systematic investing, has seen renewed interest with the application of machine learning techniques. \cite{neely2014forecasting} shows that machine learning models can identify predictable patterns in exchange rates, while \cite{gu2020empirical} demonstrates similar results for equity markets. The key insight from this literature is that successful market timing requires the integration of multiple signals across different time horizons and market regimes.

\section{Methodology}

\subsection{Algorithm Framework Architecture}

Our multi-day turnover quantitative trading algorithm is built upon a modular architecture that enables sophisticated interaction between five core components while maintaining computational efficiency and interpretability. The framework operates on a daily rebalancing schedule, processing market data through sequential modules that each contribute specialized analysis capabilities to the overall decision-making process.

The architecture follows a pipeline design where outputs from upstream modules serve as inputs to downstream components, creating a natural flow of information from raw market data to final position allocations. This design enables both parallel processing of independent computations and sequential refinement of trading signals through multiple analytical layers.

\begin{algorithm*}[t]
\caption{Multi-Day Turnover Trading Algorithm}
\begin{algorithmic}[1]
\Procedure{DailyRebalance}{$market\_data_t$}
\State $candidate\_stocks \gets \text{CrossSectionalPredict}(market\_data_t)$
\State $entry\_signals \gets \text{OpeningSignalAnalysis}(candidate\_stocks)$
\State $position\_sizes \gets \text{PositionSizing}(entry\_signals)$
\State $exit\_signals \gets \text{GridOptimizedExit}(current\_positions)$
\State $timing\_filter \gets \text{MarketTiming}(market\_data_t)$
\State $final\_positions \gets \text{IntegrateSignals}(position\_sizes, timing\_filter)$
\State \Return $final\_positions$
\EndProcedure
\end{algorithmic}
\end{algorithm*}

\subsection{Deep Cross-Sectional Prediction Module}

The cross-sectional prediction module represents the foundation of our stock selection process, employing a sophisticated deep neural network architecture specifically designed to capture relative performance patterns in the Chinese equity market. Unlike traditional factor models that rely on linear combinations of predetermined factors, our approach learns complex non-linear relationships directly from market data.

\subsubsection{Feature Engineering and Data Preprocessing}

The feature engineering process begins with the construction of a comprehensive feature set encompassing technical indicators, fundamental ratios, and market microstructure variables. Technical features include momentum indicators across multiple time horizons (5, 10, 20, 60 days), mean reversion signals, volatility measures, and volume-based indicators. Fundamental features incorporate valuation ratios, growth metrics, profitability measures, and leverage indicators.

Market microstructure features capture intraday trading patterns, including opening gap analysis, volume distribution patterns, and order flow imbalances. These features are particularly important in the A-share market context, where retail trading activity creates distinctive microstructure patterns that can be exploited for alpha generation.

Raw features undergo extensive preprocessing to ensure numerical stability and cross-sectional comparability. The normalization process applies sector-neutral standardization:

\begin{equation}
\tilde{X}_{i,t}^{(k)} = \frac{X_{i,t}^{(k)} - \mu_{sector(i),t}^{(k)}}{\sigma_{sector(i),t}^{(k)} + \epsilon}
\end{equation}

where $X_{i,t}^{(k)}$ represents feature $k$ for stock $i$ at time $t$, $sector(i)$ denotes the sector classification for stock $i$, and $\epsilon$ is a small constant to prevent division by zero.

Additionally, we apply winsorization at the 1st and 99th percentiles to control for outliers, and implement forward-fill imputation for missing values with decay weighting to avoid look-ahead bias.

\subsubsection{Neural Network Architecture}

The core prediction network employs a deep feedforward architecture with several architectural innovations designed specifically for financial applications. The network consists of multiple hidden layers with progressively decreasing dimensions, implementing an information bottleneck that forces the model to learn compressed representations of the input features.

\begin{equation}
h^{(l+1)} = \text{Dropout}(\text{ReLU}(\text{BatchNorm}(W^{(l)}h^{(l)} + b^{(l)})))
\end{equation}

Each hidden layer incorporates batch normalization to accelerate training and improve gradient flow, followed by ReLU activation and dropout regularization to prevent overfitting. The dropout rate is set to 0.3 for hidden layers and 0.1 for the input layer.

The output layer produces cross-sectional ranking scores that are normalized using a temperature-scaled softmax transformation:

\begin{equation}
P(rank_i) = \frac{\exp(z_i/T)}{\sum_{j=1}^{N} \exp(z_j/T)}
\end{equation}

where $z_i$ is the raw output for stock $i$, $T$ is the temperature parameter (set to 2.0), and $N$ is the number of stocks in the universe at time $t$.

\subsubsection{Training Methodology}

The training process employs a sophisticated approach that addresses the unique challenges of financial time series prediction. We use a walk-forward training methodology with expanding windows, where models are retrained monthly using all available historical data up to the training cutoff point.

The loss function combines ranking loss with regression objectives:

\begin{equation}
\mathcal{L} = \alpha \mathcal{L}_{ranking} + (1-\alpha) \mathcal{L}_{regression}
\end{equation}

where $\mathcal{L}_{ranking}$ is the listwise ranking loss based on future returns, and $\mathcal{L}_{regression}$ is the mean squared error between predicted and actual future returns. The weighting parameter $\alpha$ is set to 0.7, emphasizing ranking accuracy over absolute return prediction.

\subsection{Opening Signal Distribution Analysis}

The opening signal distribution analysis module exploits inefficiencies in the opening auction mechanism, which is particularly pronounced in the A-share market due to its unique trading structure and high retail participation. This component analyzes the statistical properties of overnight gaps, pre-market volume patterns, and opening price dynamics to identify profitable entry opportunities.

\subsubsection{Signal Construction}

The opening signal $S_{i,t}$ for stock $i$ at time $t$ is constructed as a weighted combination of multiple components:

\begin{equation}
S_{i,t} = \alpha_1 \cdot Gap_{i,t} + \alpha_2 \cdot VR_{i,t} + \alpha_3 \cdot Vol_{i,t} + \alpha_4 \cdot Sentiment_{i,t}
\end{equation}

where:
- $Gap_{i,t} = \frac{P_{open,t} - P_{close,t-1}}{P_{close,t-1}}$ represents the overnight gap
- $VR_{i,t} = \frac{Volume_{premarket,t}}{AVG(Volume_{t-20:t-1})}$ is the pre-market volume ratio
- $Vol_{i,t}$ captures recent volatility using GARCH(1,1) estimation
- $Sentiment_{i,t}$ incorporates market sentiment indicators

The weights $\alpha_1, \alpha_2, \alpha_3, \alpha_4$ are estimated using time-varying parameter regression with exponential decay weighting to adapt to changing market conditions.

\subsubsection{Distribution Modeling}

We model the distribution of opening signals using a mixture of three Gaussian components to capture the multi-modal nature of opening returns:

\begin{equation}
P(S_{i,t}) = \sum_{k=1}^{3} \pi_k \mathcal{N}(S_{i,t}; \mu_k, \sigma_k)
\end{equation}

The parameters $\{\pi_k, \mu_k, \sigma_k\}$ are estimated using the Expectation-Maximization algorithm with regularization to prevent overfitting:

\begin{equation}
\mathcal{L}_{EM} = -\sum_{i,t} \log P(S_{i,t}) + \lambda \sum_{k=1}^{3} |\pi_k - \frac{1}{3}|
\end{equation}

The regularization term encourages balanced mixture weights, preventing the model from collapsing to a single component.

\subsubsection{Entry Signal Generation}

Entry signals are generated when the probability of observing favorable opening conditions exceeds a dynamic threshold. The entry condition is defined as:

\begin{equation}
Entry_{i,t} = \mathbf{1}[P(S_{i,t} > \theta_t) > \phi_t \text{ and } CS_{i,t} > \psi]
\end{equation}

where $\theta_t$ and $\phi_t$ are time-varying thresholds adapted based on recent market volatility, and $CS_{i,t}$ is the cross-sectional score from the prediction module.

The dynamic threshold adaptation follows:

\begin{equation}
\theta_t = \theta_0 + \beta \cdot \text{RealizedVol}_{t-5:t-1}
\end{equation}

This ensures that entry criteria become more stringent during high volatility periods, providing natural risk control.

\subsection{Advanced Position Sizing Algorithm}

The position sizing component goes beyond traditional approaches by incorporating multiple dimensions of market structure analysis. Our algorithm simultaneously considers stock-specific characteristics, market-wide conditions, and portfolio-level constraints to optimize capital allocation.

\subsubsection{Multi-Factor Position Sizing}

The base position weight for stock $i$ is calculated using a multi-factor approach:

\begin{equation}
w_{i,base} = \frac{Score_i \cdot \sqrt{MarketCap_i} \cdot Momentum_i^{0.2}}{ADV_i^{0.3} \cdot Volatility_i^{0.5}} \cdot \lambda_i
\end{equation}

where:
- $Score_i$ is the normalized cross-sectional prediction score
- $MarketCap_i$ provides size-based weighting to improve diversification
- $Momentum_i$ incorporates recent price momentum as a quality factor
- $ADV_i$ is the 20-day average daily volume for liquidity adjustment
- $Volatility_i$ is the realized volatility for risk scaling
- $\lambda_i$ is the liquidity adjustment factor

The liquidity adjustment factor is computed to ensure position sizes remain within acceptable market impact bounds:

\begin{equation}
\lambda_i = \min\left(1, \frac{TargetVolume}{ADV_i \cdot MaxParticipation}\right)
\end{equation}

where $MaxParticipation$ is set to 10\% of average daily volume to minimize market impact.

\subsubsection{Portfolio-Level Constraints}

Position weights are subject to multiple portfolio-level constraints to ensure diversification and risk control:

\textbf{Individual Position Limits:}
\begin{equation}
0.5\% \leq w_i \leq 2.0\%
\end{equation}

\textbf{Sector Concentration Limits:}
\begin{equation}
\sum_{i \in Sector_j} w_i \leq 25\%
\end{equation}

\textbf{Market Cap Distribution:}
\begin{equation}
20\% \leq \sum_{i \in LargeCap} w_i \leq 60\%
\end{equation}

These constraints are enforced through quadratic programming optimization that minimizes tracking error to target weights while satisfying all constraints.

\subsubsection{Dynamic Adjustment Mechanism}

Position sizes are dynamically adjusted based on market conditions and portfolio performance. During high volatility periods, overall position sizes are scaled down by a factor proportional to the VIX equivalent for the Chinese market:

\begin{equation}
w_i^{adjusted} = w_i^{base} \cdot \left(1 - 0.5 \cdot \frac{VIX_{China,t} - \overline{VIX_{China}}}{\sigma_{VIX_{China}}}\right)
\end{equation}

This mechanism provides automatic risk reduction during stressed market conditions while maintaining position structure.

\subsection{Grid-Search Optimization Framework}

The grid-search optimization framework represents a systematic approach to determining optimal profit-taking and stop-loss parameters across multiple dimensions simultaneously. Rather than using fixed percentage-based rules, our system adapts parameters based on stock characteristics, market conditions, and historical performance patterns.

\subsubsection{Multi-Dimensional Parameter Space}

The optimization process explores a comprehensive parameter space defined by:

\textbf{Profit-Taking Levels:} $PT \in [1.0\%, 1.5\%, 2.0\%, 2.5\%, 3.0\%, 4.0\%, 5.0\%, 6.0\%]$

\textbf{Stop-Loss Levels:} $SL \in [0.8\%, 1.0\%, 1.2\%, 1.5\%, 2.0\%, 2.5\%, 3.0\%]$

\textbf{Maximum Holding Periods:} $MHP \in [3, 5, 7, 9, 12, 15]$ days

\textbf{Trailing Stop Activation:} $TSA \in [1.5\%, 2.0\%, 2.5\%, 3.0\%]$

This creates a 4-dimensional grid with $8 \times 7 \times 6 \times 4 = 1,344$ parameter combinations for evaluation.

\subsubsection{Multi-Objective Optimization}

The optimization objective function combines multiple performance metrics with adaptive weighting:

\begin{equation}
\begin{split}
\text{Objective} = w_1 \cdot \text{WinRate} \\
                + w_2 \cdot \frac{\text{CumReturn}}{\text{MaxDrawdown}} \\
                + w_3 \cdot \text{TurnoverEfficiency} \\
                + w_4 \cdot \text{Consistency}
\end{split}
\end{equation}

where:
- $WinRate$ is the percentage of profitable trades
- $\frac{CumReturn}{MaxDrawdown}$ represents risk-adjusted return
- $TurnoverEfficiency = \frac{AnnualReturn}{AnnualTurnover}$ measures return per unit of turnover
- $Consistency = 1 - \frac{\sigma_{MonthlyReturns}}{\mu_{MonthlyReturns}}$ captures return stability

The weights $\{w_1, w_2, w_3, w_4\}$ are set to $\{0.25, 0.35, 0.25, 0.15\}$ respectively, emphasizing risk-adjusted returns while maintaining focus on other performance dimensions.

\subsubsection{Adaptive Parameter Selection}

Rather than using fixed optimal parameters, our system implements adaptive parameter selection based on market regime identification. Market regimes are classified using a hidden Markov model with three states: Low Volatility, Normal Volatility, and High Volatility.

For each regime $r$, optimal parameters $\{PT_r, SL_r, MHP_r, TSA_r\}$ are determined through the grid-search process. The current regime is identified using the Viterbi algorithm, and parameters are adjusted accordingly with smoothing to prevent excessive switching:

\begin{equation}
Param_t = 0.7 \cdot Param_{regime,t} + 0.3 \cdot Param_{t-1}
\end{equation}

\subsection{Multi-Granularity Market Timing Model}

The market timing component integrates analysis across multiple time horizons and volatility regimes to make nuanced decisions about portfolio exposure and individual position timing. This module serves as both a risk management tool and an alpha enhancement mechanism.

\subsubsection{Multi-Scale Feature Construction}

Features for market timing are constructed across multiple time scales to capture both short-term tactical signals and medium-term strategic indicators:

\textbf{Short-term Features (1-5 days):}
- Intraday momentum and reversion patterns
- Volume-weighted price movements
- Market microstructure indicators
- Sentiment proxy variables

\textbf{Medium-term Features (5-20 days):}
- Cross-sectional dispersion measures
- Sector rotation indicators
- Volatility regime transitions
- Momentum factor performance

\textbf{Long-term Features (20-60 days):}
- Macroeconomic trend indicators
- Valuation spread metrics
- Policy environment variables
- International market correlations

\subsubsection{Hierarchical Volatility Modeling}

We implement a hierarchical approach to volatility modeling that combines multiple methodologies:

\begin{equation}
\sigma_t^2 = \alpha_1 \sigma_{GARCH,t}^2 + \alpha_2 \sigma_{RV,t}^2 + \alpha_3 \sigma_{SV,t}^2
\end{equation}

where:
- $\sigma_{GARCH,t}^2$ is the GARCH(1,1) volatility forecast
- $\sigma_{RV,t}^2$ is the realized volatility over the past 20 days  
- $\sigma_{SV,t}^2$ is the stochastic volatility estimate using particle filtering

The combination weights $\{\alpha_1, \alpha_2, \alpha_3\}$ are estimated using time-varying parameter regression with Kalman filtering to adapt to changing volatility dynamics.

\subsubsection{Regime-Aware Timing Decisions}

The timing model employs a gradient boosting framework that incorporates regime information:

\begin{equation}
F_m(x,r) = F_{m-1}(x,r) + \gamma \sum_{j=1}^{J} c_{jm}^{(r)} \mathbf{1}(x \in R_{jm}^{(r)})
\end{equation}

where $r$ represents the current market regime, and model parameters are regime-specific. This allows the timing model to adapt its behavior based on prevailing market conditions.

The final timing signal combines multiple components:

\begin{equation}
\begin{split}
TimingSignal_t = \beta_1 MomentumTiming_t \\
+ \beta_2 VolatilityTiming_t \\
+ \beta_3 SentimentTiming_t
\end{split}
\end{equation}

Each component contributes specialized information about optimal entry and exit timing across different market dimensions.

\section{Experimental Design and Implementation}

\subsection{Data Infrastructure and Universe Definition}

Our empirical analysis utilizes comprehensive A-share market data spanning from January 2010 to December 2024, providing 15 years of historical information for model training and validation. The dataset encompasses all A-share stocks listed on the Shanghai and Shenzhen exchanges, with careful attention to survivorship bias elimination through the inclusion of delisted securities and suspension periods.

The investment universe is dynamically defined using multiple filters to ensure tradability and data quality. Primary filters include: (1) Minimum market capitalization of 500 million RMB to ensure adequate liquidity; (2) Minimum average daily trading volume of 10 million RMB over the past 20 trading days; (3) Exclusion of stocks with special treatment (ST) designations or those under suspension; (4) Minimum trading history of 252 days (one trading year) for newly listed stocks; (5) Exclusion of stocks with extreme price movements (>30

The feature set comprises over 200 variables across multiple categories. Technical indicators include various momentum measures (RSI, MACD, Williams 

\subsection{Training and Validation Methodology}

The model training process follows a rigorous walk-forward methodology designed to simulate realistic trading conditions while avoiding look-ahead bias. The training framework divides the historical period into three distinct phases:

\textbf{Training Period:} 2010-2020 (10 years) - Used for initial model training and hyperparameter optimization across all components.

\textbf{Validation Period:} 2020-2021 (1 year) - Used for model selection, parameter tuning, and strategy component integration.

\textbf{Out-of-Sample Testing:} 2021-2024 (4 years) - Pure out-of-sample evaluation with no parameter adjustments.

Within the training period, we employ expanding window retraining where models are updated monthly using all available historical data up to the retraining date. This approach ensures that models incorporate the most recent market dynamics while maintaining sufficient historical context for robust parameter estimation.

Cross-validation is implemented using time-series aware techniques that respect temporal ordering. We use blocked cross-validation with 6-month training blocks and 1-month validation blocks, ensuring that validation periods never precede training periods. Hyperparameter optimization is performed using Bayesian optimization with Gaussian process priors to efficiently explore the parameter space.

\subsection{Performance Measurement Framework}

Our performance evaluation framework employs both traditional and advanced metrics to comprehensively assess strategy performance across multiple dimensions. Standard metrics include annualized returns, volatility, Sharpe ratio, maximum drawdown, and various risk-adjusted performance measures.

Advanced metrics capture aspects particularly relevant to institutional implementation: Information Ratio relative to sector-neutral benchmarks, Calmar Ratio (return/maximum drawdown), Sortino Ratio using downside deviation, Omega Ratio for higher-moment risk assessment, and various tail risk measures including Expected Shortfall and Maximum Daily Loss.

Transaction cost modeling incorporates realistic estimates based on A-share market structure. We assume 0.05

\begin{equation}
MarketImpact = 0.5 \cdot \sqrt{\frac{SharesTraded}{ADV}} \cdot Volatility \cdot Sign
\end{equation}

This formulation captures the non-linear relationship between trade size and market impact while accounting for volatility effects.

\section{Results and Comprehensive Analysis}

\subsection{Overall Performance Metrics}

Table \ref{tab:performance_detailed} presents comprehensive performance statistics for our algorithm compared to various benchmarks over the out-of-sample testing period (2021-2024).

\begin{table*}[htbp]
\caption{Detailed Performance Comparison (2021-2024)}
\begin{center}
\begin{tabular}{|l|c|c|c|c|c|c|}
\hline
\textbf{Metric} & \textbf{Our Algorithm} & \textbf{CSI 300} & \textbf{CSI 500} & \textbf{Equal Weight} & \textbf{Market Neutral} & \textbf{Factor Model} \\
\hline
Annualized Return & 15.2\% & 2.8\% & 4.1\% & 5.1\% & 8.7\% & 11.3\% \\
Annualized Volatility & 8.1\% & 18.3\% & 21.2\% & 16.4\% & 9.2\% & 12.1\% \\
Maximum Drawdown & 4.8\% & 28.4\% & 31.7\% & 22.3\% & 7.2\% & 9.8\% \\
Sharpe Ratio & 1.87 & 0.15 & 0.19 & 0.31 & 0.95 & 0.93 \\
Information Ratio & 1.12 & - & - & 0.43 & 0.67 & 0.81 \\
Calmar Ratio & 3.17 & 0.10 & 0.13 & 0.23 & 1.21 & 1.15 \\
Sortino Ratio & 2.84 & 0.21 & 0.26 & 0.45 & 1.38 & 1.42 \\
Win Rate & 58.3\% & - & - & 51.2\% & 54.1\% & 55.7\% \\
Average Holding Period & 6.2 days & - & - & 9.0 days & 8.5 days & 7.8 days \\
Annual Turnover & 2100\% & - & - & 1800\% & 1650\% & 1950\% \\
Maximum Daily Loss & -1.2\% & -8.7\% & -9.3\% & -6.4\% & -2.1\% & -2.8\% \\
VaR (95\%) & -0.8\% & -3.2\% & -3.7\% & -2.6\% & -1.1\% & -1.5\% \\
Expected Shortfall & -1.1\% & -5.1\% & -5.8\% & -4.2\% & -1.6\% & -2.1\% \\
\hline
\end{tabular}
\label{tab:performance_detailed}
\end{center}
\end{table*}

Our algorithm demonstrates exceptional performance across multiple dimensions, achieving superior risk-adjusted returns while maintaining low volatility and drawdown characteristics. The Sharpe ratio of 1.87 represents substantial outperformance relative to both market benchmarks and alternative systematic strategies.

Particularly noteworthy is the algorithm's ability to maintain consistent performance during stressed market periods. The maximum daily loss of -1.2\% compares favorably to the -8.7\% experienced by the CSI 300 during the same period, demonstrating the effectiveness of our integrated risk management approach.

\subsection{Component Contribution Analysis}

We conduct detailed ablation studies to quantify the contribution of each algorithmic component to overall performance. Table \ref{tab:ablation} presents the incremental impact of each module.

\begin{table}[htbp]
\caption{Component Ablation Analysis}
\begin{center}
\begin{tabular}{|l|c|c|c|c|}
\hline
\textbf{Configuration} & \textbf{Return} & \textbf{Sharpe} & \textbf{MaxDD} & \textbf{Win Rate} \\
\hline
Baseline (Random) & 3.2\% & 0.18 & 15.2\% & 49.1\% \\
+ Cross-Sectional & 11.4\% & 0.87 & 8.9\% & 54.2\% \\
+ Opening Signals & 13.1\% & 1.12 & 7.6\% & 56.1\% \\
+ Position Sizing & 14.3\% & 1.34 & 6.2\% & 57.3\% \\
+ Grid Optimization & 14.8\% & 1.61 & 5.1\% & 58.9\% \\
+ Market Timing & 15.2\% & 1.87 & 4.8\% & 58.3\% \\
\hline
\end{tabular}
\label{tab:ablation}
\end{center}
\end{table}

The cross-sectional prediction module contributes the largest individual impact, adding 8.2 percentage points to annual returns and significantly improving the win rate. The opening signal analysis provides additional timing precision, contributing 1.7 percentage points while further reducing drawdown. The sophisticated position sizing algorithm contributes 1.2 percentage points while substantially reducing volatility. Grid optimization adds 0.5 percentage points while materially improving the Sharpe ratio through enhanced risk management. Finally, the market timing component contributes 0.4 percentage points while providing the final reduction in maximum drawdown.

\subsection{Regime-Based Performance Analysis}

We analyze performance across different market regimes to assess strategy robustness. Market regimes are defined based on volatility levels and trend directions:

\textbf{Bull Market (Low Vol):} VIX < 20th percentile, positive 3-month trend
\textbf{Bull Market (High Vol):} VIX > 80th percentile, positive 3-month trend  
\textbf{Bear Market (Low Vol):} VIX < 20th percentile, negative 3-month trend
\textbf{Bear Market (High Vol):} VIX > 80th percentile, negative 3-month trend
\textbf{Sideways Market:} All other periods

\begin{table}[htbp]
\caption{Regime-Based Performance Analysis}
\begin{center}
\begin{tabular}{|l|c|c|c|c|c|}
\hline
\textbf{Regime} & \textbf{Days} & \textbf{Return} & \textbf{Sharpe} & \textbf{MaxDD} & \textbf{Win Rate} \\
\hline
Bull/Low Vol & 156 & 22.3\% & 2.41 & 2.1\% & 64.2\% \\
Bull/High Vol & 89 & 18.7\% & 1.52 & 3.8\% & 59.1\% \\
Bear/Low Vol & 134 & 8.9\% & 1.23 & 3.2\% & 52.4\% \\
Bear/High Vol & 142 & 4.2\% & 0.87 & 4.8\% & 49.7\% \\
Sideways & 445 & 12.1\% & 1.67 & 2.9\% & 57.8\% \\
\hline
\end{tabular}
\label{tab:regime_performance}
\end{center}
\end{table}

The algorithm demonstrates robust performance across all market regimes, with particularly strong results during favorable conditions while maintaining reasonable performance during challenging periods. Even during bear markets with high volatility, the strategy maintains positive returns with controlled drawdowns.

\subsection{Sector and Style Analysis}

Our strategy's performance is further analyzed across different market sectors and investment styles to understand sources of alpha generation.

\begin{table*}[htbp]
\caption{Sector Exposure and Performance}
\begin{center}
\begin{tabular}{|l|c|c|c|c|}
\hline
\textbf{Sector} & \textbf{Avg Weight} & \textbf{Contribution} & \textbf{Hit Rate} & \textbf{Avg Return} \\
\hline
Technology & 18.2\% & 3.1\% & 61.4\% & 17.0\% \\
Healthcare & 15.8\% & 2.7\% & 59.8\% & 17.1\% \\
Consumer Disc. & 14.6\% & 2.3\% & 58.2\% & 15.8\% \\
Industrials & 13.9\% & 2.1\% & 57.1\% & 15.1\% \\
Materials & 12.1\% & 1.8\% & 56.3\% & 14.9\% \\
Consumer Staples & 10.7\% & 1.6\% & 55.7\% & 14.9\% \\
Financials & 8.2\% & 1.0\% & 53.4\% & 12.2\% \\
Energy & 6.5\% & 0.6\% & 51.9\% & 9.2\% \\
\hline
\end{tabular}
\label{tab:sector_analysis}
\end{center}
\end{table*}

The algorithm shows balanced sector exposure with a slight tilt toward growth sectors (Technology and Healthcare), which aligns with the superior performance observed in these areas. The hit rates across sectors demonstrate consistent alpha generation capabilities.

\subsection{Risk Decomposition Analysis}

We perform comprehensive risk decomposition to understand the sources of portfolio risk and validate our risk management approach.

\begin{table}[htbp]
\caption{Risk Factor Decomposition}
\begin{center}
\begin{tabular}{|l|c|c|c|}
\hline
\textbf{Risk Factor} & \textbf{Exposure} & \textbf{Contribution to Risk} & \textbf{t-statistic} \\
\hline
Market Beta & 0.12 & 15.2\% & 2.1 \\
Size Factor & -0.08 & 8.3\% & -1.4 \\
Value Factor & 0.05 & 3.2\% & 0.8 \\
Momentum Factor & 0.23 & 22.1\% & 3.7 \\
Quality Factor & 0.18 & 12.4\% & 2.9 \\
Volatility Factor & -0.14 & 9.8\% & -2.2 \\
Liquidity Factor & -0.06 & 4.1\% & -1.1 \\
Idiosyncratic Risk & - & 24.9\% & - \\
\hline
\end{tabular}
\label{tab:risk_decomposition}
\end{center}
\end{table}

The risk decomposition reveals well-controlled factor exposures with moderate momentum and quality tilts that contribute positively to performance. The low market beta (0.12) confirms the strategy's market-neutral characteristics, while the substantial idiosyncratic risk component (24.9\%) indicates effective stock-specific alpha generation.

\subsection{Transaction Cost and Capacity Analysis}

Understanding transaction costs and strategy capacity is crucial for institutional implementation. We model transaction costs comprehensively and analyze capacity constraints across multiple dimensions.

\begin{table*}[htbp]
\caption{Transaction Cost Breakdown}
\begin{center}
\begin{tabular}{|l|c|c|c|}
\hline
\textbf{Cost Component} & \textbf{Basis Points} & \textbf{Annual Impact} & \textbf{Percentage of Gross Return} \\
\hline
Commission & 5.0 & 105 bp & 6.9\% \\
Stamp Tax & 10.0 & 210 bp & 13.8\% \\
Market Impact & 3.2 & 67 bp & 4.4\% \\
Bid-Ask Spread & 2.1 & 44 bp & 2.9\% \\
Timing Cost & 1.8 & 38 bp & 2.5\% \\
\textbf{Total} & \textbf{22.1} & \textbf{464 bp} & \textbf{30.5\%} \\
\hline
\end{tabular}
\label{tab:transaction_costs}
\end{center}
\end{table*}

Transaction costs represent a significant but manageable component of gross returns. The strategy's emphasis on position sizing based on liquidity constraints helps control market impact costs, while the multi-day holding period reduces turnover relative to day-trading strategies.

Capacity analysis indicates the strategy can accommodate substantial assets under management. Conservative estimates suggest capacity of 8-12 billion RMB while maintaining performance characteristics, with potential for higher capacity through careful implementation and market impact modeling.

\subsection{Stress Testing and Scenario Analysis}

We conduct extensive stress testing to evaluate strategy resilience under various adverse scenarios.

\begin{table*}[htbp]
\caption{Stress Test Results}
\begin{center}
\begin{tabular}{|l|c|c|c|c|}
\hline
\textbf{Scenario} & \textbf{Duration} & \textbf{Strategy Return} & \textbf{Maximum DD} & \textbf{Recovery Time} \\
\hline
2015 A-Share Crisis & 3 months & -2.1\% & 3.8\% & 2.1 months \\
2018 Trade Tensions & 6 months & +1.7\% & 2.9\% & N/A \\
2020 COVID Pandemic & 4 months & +3.2\% & 2.1\% & N/A \\
2022 Lockdown Period & 2 months & -0.8\% & 1.9\% & 1.2 months \\
Hypothetical 30\% Market Decline & N/A & -4.5\% & 5.2\% & 3.5 months \\
\hline
\end{tabular}
\label{tab:stress_tests}
\end{center}
\end{table*}

The stress testing results demonstrate exceptional resilience across various crisis scenarios. During the severe 2015 A-share market crisis, our strategy experienced only -2.1\% returns compared to -45\% for the broader market. The rapid recovery times and controlled drawdowns validate the effectiveness of our integrated risk management approach.

\section{Discussion and Implications}

\subsection{Strategic Advantages and Innovation}

Our multi-day turnover quantitative trading algorithm represents several significant innovations in systematic trading methodology. The integration of deep learning techniques with traditional quantitative methods creates a robust framework that captures both linear and non-linear relationships in market data while maintaining interpretability and risk control.

The cross-sectional prediction approach addresses a fundamental challenge in equity selection by focusing on relative rather than absolute performance prediction. This methodology naturally hedges market risk while concentrating on stock-specific alpha generation, resulting in more consistent risk-adjusted returns across varying market conditions.

The opening signal distribution analysis component exploits a unique feature of the A-share market structure that has not been systematically addressed in previous literature. By modeling the statistical properties of opening auctions and incorporating this information into entry timing decisions, we create an additional source of alpha that is particularly well-suited to the Chinese market context.

The adaptive position sizing algorithm represents a significant advancement over traditional approaches by incorporating real-time liquidity and market impact considerations. This dynamic approach enables the strategy to maintain performance characteristics across different market regimes while scaling to institutional asset levels.

The grid-search optimization framework for exit parameters moves beyond static rule-based approaches to create a comprehensive optimization process that balances multiple performance objectives. This methodology ensures that risk management parameters adapt to changing market conditions while maintaining focus on long-term performance consistency.

\subsection{Risk Management Framework}

The integrated risk management approach embedded throughout our algorithm represents a comprehensive framework for controlling multiple dimensions of portfolio risk. Unlike traditional strategies that rely primarily on position-level risk controls, our approach incorporates risk management at every stage of the investment process.

The cross-sectional approach naturally provides market neutrality, reducing systematic risk exposure while maintaining focus on idiosyncratic alpha generation. The opening signal analysis includes built-in volatility adjustments that automatically reduce position sizes during stressed market conditions. The position sizing algorithm incorporates explicit liquidity constraints that prevent excessive market impact while maintaining diversification benefits.

The grid-optimization framework provides dynamic adjustment of exit parameters based on market regime identification, ensuring that risk controls remain appropriate across different market environments. The market timing component serves as a final risk overlay that can reduce overall portfolio exposure during particularly challenging market conditions.

This multi-layered approach to risk management has proven effective across various market stress scenarios, as demonstrated in our comprehensive stress testing analysis. The strategy's ability to maintain positive returns during adverse market conditions while limiting drawdowns validates the effectiveness of this integrated approach.

\subsection{Scalability and Implementation Considerations}

The strategy's design explicitly addresses scalability concerns that limit many systematic trading approaches. The focus on liquid, large-capitalization stocks combined with position sizing constraints based on average daily volume ensures that the strategy can accommodate substantial assets under management without significant performance degradation.

The 9-day maximum holding period balances the need for sufficient time for alpha realization with capital efficiency considerations. This holding period is long enough to allow fundamental value to be recognized by the market while short enough to maintain high capital turnover and responsiveness to changing market conditions.

The strategy's capacity for 50-100 daily positions provides substantial diversification benefits while remaining within the operational capabilities of institutional trading desks. The position sizes of 0.5-2.0

Implementation considerations include the need for robust data infrastructure, sophisticated order management systems, and experienced portfolio management teams. The strategy's complexity requires careful monitoring and periodic recalibration, but the modular design enables component-level analysis and adjustment without disrupting overall performance.

\subsection{Market Structure Implications}

Our results have important implications for understanding alpha generation opportunities in the Chinese A-share market. The superior performance of the opening signal analysis component suggests that market microstructure inefficiencies remain exploitable despite increasing market sophistication.

The effectiveness of cross-sectional prediction models indicates that fundamental analysis combined with advanced machine learning techniques can identify persistent mispricings in individual securities. This finding contradicts strong-form market efficiency hypotheses and suggests continued opportunities for systematic alpha generation.

The regime-dependent performance characteristics observed in our analysis highlight the importance of adaptive strategy design in emerging markets. The ability to maintain positive returns across different market environments while adjusting risk characteristics appropriately represents a significant advancement in systematic trading methodology.

\subsection{Limitations and Future Research Directions}

Despite the strong empirical results, several limitations warrant acknowledgment. The strategy's performance is inherently dependent on continued market inefficiencies that may diminish as similar methodologies become more widespread. The complexity of the integrated approach requires substantial infrastructure investment and ongoing maintenance.

The four-year out-of-sample testing period, while substantial, may not capture all possible market regimes or structural changes that could affect future performance. The strategy's reliance on historical relationships assumes some degree of market structure stability that may not persist indefinitely.

Future research directions include the integration of alternative data sources such as satellite imagery, social media sentiment, and corporate communications analysis. The application of reinforcement learning techniques to dynamic strategy adaptation represents another promising avenue for enhancement.

Cross-market extensions to Hong Kong H-shares and international emerging markets could provide additional diversification benefits while testing the generalizability of our methodological innovations. The development of real-time strategy optimization using online learning techniques could further enhance performance consistency.

Advanced risk modeling incorporating tail risk dependencies and regime-switching models could provide additional robustness during extreme market conditions. The integration of ESG factors and sustainable investing considerations represents an important area for future development.

\section{Conclusion}

This research presents a comprehensive multi-day turnover quantitative trading algorithm that successfully integrates advanced deep learning techniques with sophisticated risk management methodologies to generate superior risk-adjusted returns in the Chinese A-share market. Our approach addresses fundamental challenges in systematic trading through innovative solutions that balance alpha generation with risk control and scalability considerations.

The empirical results demonstrate exceptional performance characteristics with 15.2\% annualized returns, 4.8\% maximum drawdown, and 1.87 Sharpe ratio over a four-year out-of-sample testing period. These results represent substantial outperformance relative to both market benchmarks and alternative systematic strategies while maintaining characteristics suitable for institutional implementation.

The key innovations of our approach include the integration of cross-sectional prediction with opening signal distribution analysis, the development of adaptive position sizing algorithms incorporating real-time liquidity constraints, the implementation of comprehensive grid-search optimization for dynamic risk management, and the creation of multi-granularity market timing models that operate across multiple time horizons.

The strategy's robust performance across different market regimes, sectors, and stress scenarios validates the effectiveness of our integrated risk management approach. The comprehensive analysis of transaction costs, capacity constraints, and implementation considerations demonstrates the practical viability of our methodology for institutional deployment.

The slightly lagged but stable entry timing characteristic, while potentially missing some immediate opportunities, provides significant risk mitigation benefits that align well with institutional risk management requirements. This conservative approach enables the strategy to avoid major drawdowns while maintaining attractive return profiles across varying market conditions.

Our work contributes to the growing literature on machine learning applications in quantitative finance while addressing practical implementation challenges that limit the real-world applicability of many academic approaches. The modular design enables continued enhancement and adaptation as market conditions evolve and new data sources become available.

The strategy's high capacity for asset management, demonstrated ability to generate consistent alpha across market cycles, and robust risk management framework make it well-suited for institutional implementation. The comprehensive performance analysis and stress testing provide confidence in the strategy's resilience under adverse market conditions.

Future research will focus on incorporating alternative data sources, developing more sophisticated adaptive mechanisms, and extending the methodology to additional markets. The integration of reinforcement learning techniques and real-time optimization represents particularly promising directions for further enhancement.

This work demonstrates that sophisticated integration of modern machine learning techniques with traditional quantitative methods can create systematic trading strategies that generate superior risk-adjusted returns while maintaining the robustness and scalability required for institutional implementation. The success of our approach in the challenging Chinese A-share market environment suggests broad applicability to other emerging market contexts.

\section*{Acknowledgments}

The authors thank the anonymous reviewers for their constructive feedback and suggestions that significantly improved the quality and clarity of this paper. We also acknowledge the valuable discussions with practitioners and academics that helped shape our understanding of the practical implementation challenges addressed in this work.

\end{document}